\documentclass[useAMS,usenatbib]{mn2e}
\usepackage{graphicx,amssym}
\newcommand{\bc}{\begin{center}}
\newcommand{\ec}{\end{center}}

\title[The environmental history of group and cluster galaxies]
      {The environmental history of group and cluster galaxies in a 
        $\Lambda$CDM Universe}  
\author[G.~De Lucia et al.]
       {Gabriella De Lucia$^{1}$\thanks{Email: delucia@oats.inaf.it}, 
        Simone Weinmann$^{2}$, Bianca M. Poggianti$^3$, Alfonso \newauthor 
        Arag\'on-Salamanca$^4$, Dennis Zaritsky$^5$\\
        $^1$INAF - Astronomical Observatory of Trieste, via G.B. Tiepolo 11, 
        I-34143 Trieste, Italy\\
        $^2$Leiden Observatory, Leiden University, PO Box 9513, 2300 RA Leiden,
        the Netherlands\\ 
        $^3$INAF - Astronomical Observatory of Padova, Vicolo dell'Osservatorio
	5, Padova I-35122, Italy \\
        $^4$School of Physics and Astronomy, The University of Nottingham,
        University Park, Nottingham NG7 2RD\\ 
        $^5$Steward Observatory, University of Arizona, Tucson, AZ 85721, USA}
\begin{document}

\pagerange{\pageref{firstpage}--\pageref{lastpage}} 
\pubyear{2011}

\maketitle

\label{firstpage}

\begin{abstract}
We use publicly available galaxy merger trees, obtained applying semi-analytic
techniques to a large high resolution cosmological simulation, to study the
environmental history of group and cluster galaxies. Our results highlight the
existence of an intrinsic {\it history bias} which makes the nature versus
nurture (as well as the mass versus environment) debate inherently ill posed.
In particular we show that: (i) surviving massive satellites were accreted
later than their less massive counterparts, from more massive haloes; (ii) the
mixing of galaxy populations is incomplete during halo assembly, which creates
a correlation between the time a galaxy becomes satellite and its present
distance from the parent halo centre. The weakest trends are found for the most
massive satellites, as a result of efficient dynamical friction and late
formation times of massive haloes. A large fraction of the most massive
group/cluster members are accreted onto the main progenitor of the final halo
as central galaxies, while about half of the galaxies with low and intermediate
stellar mass are accreted as satellites. Large fractions of group and cluster
galaxies (in particular those of low stellar mass) have therefore been
`pre-processed' as satellites of groups with mass $\sim 10^{13}\,{\rm
  M}_{\odot}$. To quantify the relevance of hierarchical structure growth on
the observed environmental trends, we have considered observational estimates
of the passive galaxy fractions, and their variation as a function of halo mass
and cluster-centric distance. Comparisons with our theoretical predictions
require relatively long times ($\sim 5-7$~Gyr) for the suppression of star
formation in group and cluster satellites. It is unclear how such a gentle mode
of strangulation can be achieved by simply relaxing the assumption of
instantaneous stripping of the hot gas reservoir associated with accreting
galaxies, or if the difficulties encountered by recent galaxy formation models
in reproducing the observed trends signal a more fundamental problem with the
treatment of star formation and feedback in these galaxies.
\end{abstract}


\section{Introduction}
\label{sec:intro}

The observed properties of galaxies have long been known to depend on the
`environment' in which they are located. This correlation was quantified by
early work showing that galaxy clusters are distinguished by lower fractions of
star forming, disc dominated galaxies than regions of `average' density
\citep[e.g.][]{Hubble_and_Humason_1931,Oemler_1974,Dressler_1980}. In recent
years, observational studies trying to assess the role of environment on galaxy
evolution have received much impetus from the completion of large spectroscopic
and photometric surveys at different cosmic epochs \citep[e.g.][ just to
  mention a
  few]{Kauffmann_etal_2004,Balogh_etal_2004,Cucciati_etal_2006,Cooper_etal_2006}.
Despite much effort, however, disentangling the environment(s) and related
physical processes that are responsible for the observed trends has proved
difficult, and their physical origin is still subject of an active debate.

Much of the argument centres on whether these trends are the end product of
physical processes coming into play only after galaxies have become part of a
`group' or of a `cluster' (the {\it nurture} hypothesis), or whether they are
established before these events take place, due to galaxy formation proceeding
differently in overdense regions (the {\it nature} hypothesis). In the current
standard paradigm for structure formation, dark matter collapses into haloes in
a bottom-up fashion: small objects form first and subsequently merge into
progressively larger systems. As structure grows, galaxies join more and more
massive systems, therefore experiencing a variety of environments during their
lifetimes. In this context, the {\it nature-nurture debate appears to be ill
  posed}, as these two elements of galaxy evolution are inevitably and heavily
intertwined.

As for the nurture scenario, a variety of physical processes might be effective
in suppressing star formation and affecting the morphology of cluster galaxies.
Broadly speaking, these can be grouped in two big families: (i) interactions
with other cluster members and/or with the cluster potential, and (ii)
interactions with the hot gas that permeates massive galaxy systems. The
influence of these physical processes and their characteristic time-scales have
been studied using detailed numerical simulations \citep[for a review, see
  e.g.][]{DeLucia_2010}. This work has shown, for example, that the pressure
experienced by galaxies travelling through a dense intra-cluster medium ({\it
  ram-pressure}) can be effective in removing the galaxy interstellar-medium,
thereby suppressing subsequent episodes of star formation
\citep{Gunn_and_Gott_1972}. Ram-pressure is expected to be more important at
the centre of massive clusters, because of the large relative velocities and
higher densities of the intra-cluster medium. Galaxies orbiting in massive
clusters also experience repeated fast encounters with other cluster
members. The cumulative effect of these encounters ({\it harassment}) can drive
a strong internal dynamical response, leading for example to the transformation
of spiral galaxies into dwarf spheroidals
\citep{Farouki_and_Shapiro_1981,Moore_etal_1998}.

As discussed above, however, these physical processes should be coupled with a
{\it history bias} that is integral part of the hierarchical structure
formation. Until about one decade ago, this effect was believed to play a minor
role: early numerical work found no dependence of the clustering of dark matter
haloes on their properties, such as concentration or formation
time\footnote{This is usually defined as the time when half of the final mass
  of the halo is first assembled in a single object.}
\citep{Lemson_and_Kauffmann_1999,Percival_etal_2003}. Taking advantage of high
resolution simulations of structure formation, however, recent studies have
demonstrated that halo properties like concentration, spin, shape, and internal
angular momentum exhibit clear environmental dependencies
\citep[e.g.][]{Avila-Reese_etal_2005}. Haloes in {\it over-dense} regions form
statistically earlier and merge more rapidly than haloes in regions of the
Universe of {\it average} density
\citep[][]{Gao_etal_2005,Maulbetsch_etal_2007}. This differential evolution is
bound to leave an `imprint' on the observable properties of galaxies that
inhabit different regions at any cosmic epoch. In this context, a crucial
missing ingredient for a correct interpretation of the observed environmental
trends is represented by a detailed characterization and quantification of the
{\it environmental history} of group and cluster galaxies.

A few recent studies have touched this issue. \citet{Brueggen_and_DeLucia_2008}
combined semi-analytic models of galaxy formation with analytic models for the
gas distribution in clusters to study the ram-pressure histories of present day
cluster galaxies. They showed that virtually all cluster galaxies suffered
episodes of ram-pressure during their life-time, and argued that this physical
process might have a significant role in shaping the observed properties of the
entire cluster galaxy population. More recently, \citet{Berrier_etal_2009} and
\citet{McGee_etal_2009} have studied the accretion history of galaxies onto
clusters with the aim to quantify the relevance of pre-processing in galaxy
groups. These two studies make use of different methods, and reach different
conclusions. In the following, we will discuss findings from these studies in
more detail, and will compare them with our results.

In this study, we adopt an approach similar to that employed by
\citet{McGee_etal_2009}. In particular, we take advantage of publicly available
galaxy merger trees, obtained by applying semi-analytic techniques to a large
high resolution cosmological simulation. These merger trees are analysed in
order to study the history of the environments that galaxies have experienced
during their lifetime, as a function of the parent halo mass at present day,
and as a function of present day galaxy stellar mass.  The layout of the paper
is as follows: in Section~\ref{sec:simsam}, we provide a brief description of
the models used in our study, and describe the method and definitions
adopted. In Section~\ref{sec:history}, we characterize the history of galaxies
in terms of their parent halo mass, and define times that should play a
relevant role in their evolution. In Section~\ref{sec:bias}, we discuss the
expected trends due to the environmental history for galaxies of different
stellar mass. We also compare these expectations with observational estimates
of the fraction of red and passive galaxies as a function of parent halo mass,
galaxy stellar mass, and cluster-centric distance. Finally, we discuss our
results in Section~\ref{sec:disc}, and summarize our conclusions in
Section~\ref{sec:concl}.

\section{The galaxy formation models}
\label{sec:simsam}

In this study, we take advantage of the publicly available catalogues from the
galaxy formation model presented in \citet{DeLucia_and_Blaizot_2007}, and
applied to the Millennium Simulation \citep{Springel_etal_2005}. This
simulation follows $N=2160^3$ particles of mass $8.6\times10^{8}\,h^{-1}{\rm
  M}_{\odot}$ within a comoving box of $500\,h^{-1}$Mpc on a side. It is based
on a $\Lambda$CDM model with parameters $\Omega_{\rm m}=0.25$, $\Omega_{\rm
  b}=0.045$, $h=0.73$, $\Omega_\Lambda=0.75$, $n=1$, and $\sigma_8=0.9$, where
the Hubble constant is parameterised as $H_0 = 100\, h\, {\rm km\, s^{-1}
  Mpc^{-1}}$. The simulation outputs were used to construct merger trees of all
gravitationally self-bound dark matter subhaloes down to $20$ particles, which
corresponds to a mass of $1.7\times10^{10}\,h^{-1} M_{\sun}$. These merger
trees represent the basic input needed for the semi-analytic model described in
\citet{DeLucia_and_Blaizot_2007}.

Down to the resolution limit of the Millennium Simulation, this model has been
shown to provide a reasonable agreement with a large variety of observational
data, both in the local Universe and at high redshift. It is, however, not
without problems. In particular, the predicted galaxy stellar mass function
exhibits an excess of low-to-intermediate mass galaxies at all redshifts, the
fraction of red galaxies among low mass galaxies is too high compared to
observational data, and the model over-predicts the clustering signal of faint
galaxies
\citep[e.g.][]{Weinmann_etal_2006,Wang_etal_2008,Fontanot_etal_2009}. It is
important to note that these `failures' are not specific of the particular
model adopted in this study. Rather, they appear to be common to most (all)
models that have been published recently, and have likely related causes. No
satisfactory explanation and/or solution has been found yet to solve all of the
problems mentioned above (for recent attempts, see e.g. \citealt{Guo_etal_2011}
and \citealt*{Wang_etal_2011}).

The model used in this study neglects environmental physical processes such as
ram-pressure and harassment, but assumes that when galaxies are accreted onto a
more massive system, the associated hot gas reservoir is stripped
instantaneously. This induces a very rapid decline of the star formation
histories of satellite galaxies, and contributes to create an excess of red and
passive galaxies with respect to the observations
\citep[e.g.][]{Wang_etal_2007}. In recent studies, a more gradual stripping of
the hot gas reservoir has been assumed
\citep[][]{Kang_and_vandenBosch_2008,Font_etal_2008,Weinmann_etal_2010,Guo_etal_2011},
following results from numerical simulations \citep[][but see also
  \citealt{Saro_etal_2010}]{McCarthy_etal_2008}. Albeit improved, the agreement
with observational measurements is far from satisfactory
\citep[e.g.][]{Guo_etal_2011,Weinmann_etal_2011}.  In this study, we will make
limited use of the physical properties of galaxies as predicted by the
model. In particular, we will use only the predicted galaxy stellar mass, and
we will focus on the resulting galaxy merger trees and their dependence on the
parent halo mass. In the following, we will focus on haloes selected from the
box of the simulation corresponding to z=0, with mass larger than $\sim
10^{13}\,{\rm M}_{\odot}$. The halo mass is computed from the $N$-body
simulation, as the mass within a sphere enclosing a mean overdensity that is
equal to 200 times the critical density of the Universe ($M_{200}$). At the
resolution of the Millennium Simulation, haloes with $M_{200} \sim
10^{13}\,{\rm M}_{\odot}$ contain on average about $15$ galaxies within the
virial radius ($R_{200}$ in this study), with stellar mass larger than $\gtrsim
10^9\,{\rm M_{\odot}}$. To investigate trends as a function of halo mass, we
have randomly selected fifteen haloes in each of the following mass bins: $\sim
10^{13}$, $\sim 5\times 10^{13}$, $10^{14}$, and $\sim 5\times 10^{14}\,{\rm
  M}_{\odot}$. These haloes contain 9201 satellite galaxies within their virial
radii, that represent the basic sample of our analysis. 6929 of these
satellites have present day stellar mass between $10^9$ and $10^{10}\,{\rm
  M}_{\sun}$, 2081 have mass between $10^{10}$ and $10^{11}\,{\rm M}_{\sun}$,
and only 191 of them have stellar mass larger than $10^{11}\,{\rm M}_{\sun}$.

\begin{figure}
\bc
\resizebox{8.5cm}{!}{\includegraphics[angle=90]{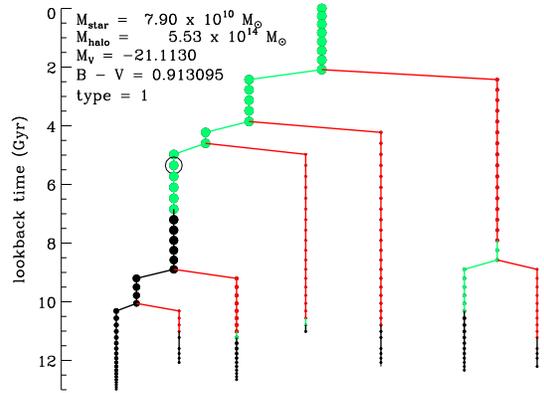}} 
\caption{Example merger tree for a galaxy of stellar mass $\sim 8\times
  10^{10}\,{\rm M}_{\odot}$, residing in a halo of mass $\sim
  5.5\times10^{14}\,{\rm M}_{\odot}$. The size of the symbols scales with the
  galaxy stellar mass while different colours are used for central galaxies
  (black), satellite galaxies associated with distinct dark matter subhaloes
  (green), and satellites whose parent substructures were stripped below the
  resolution of the simulation (red). The empty black circle at lookback time
  $\sim 5.5$~Gyr corresponds to the lookback time before the galaxy is accreted
  onto the main progenitor of the final halo.}
\label{fig:trexample}
\ec
\end{figure}

In order to clarify the method and the definitions that we have adopted and
that we will use in the following, we show in Figure~\ref{fig:trexample} an
example of a galaxy merger tree. The galaxy considered is shown at the top of
the plot, and all its progenitors (and their histories) are shown going
backward in time. The size of the symbols scales with the galaxy stellar mass,
while different colours are used for central galaxies (shown in black),
satellite galaxies associated with a distinct dark matter substructure
(green), and satellite galaxies whose parent dark matter subhaloes have been
stripped below the resolution of the simulation (these are shown in red). We
note that only central galaxies are fuelled by cooling flows in our model. The
leftmost branch in Figure~\ref{fig:trexample} is obtained by connecting the
galaxy to its most massive progenitor (often referred to as the {\it main
  progenitor}), at each node of the tree.

In the following, when tracing the history of our model galaxies, we will
effectively refer to the main progenitor of the galaxy at each time. In
addition, we define two different {\it characteristic times}: (i) the lookback
time before a galaxy becomes satellite of a larger halo, and (ii) the lookback
time before the galaxy is accreted onto the main progenitor of the final
halo. In the following, we will often refer to the former characteristic time
as $t_{\rm sat}$ or the time when {\it the galaxy becomes a satellite}, and to
the latter as $t_{\rm halo}$ or the time when {\it the galaxy is accreted onto
  the final group/cluster}. In the example shown in Figure~\ref{fig:trexample},
$t_{\rm sat}$ corresponds to the last time when the colour of the main
progenitor is black before becoming green ($\sim 7$~Gyr), while $t_{\rm halo}$
is marked by an empty black symbol ($\sim 5.5$~Gyr for the example
considered). As we will discuss in detail in the following, the adoption of
different characteristic times is largely responsible for the different
conclusions reached by \citet{Berrier_etal_2009} and \citet{McGee_etal_2009}.

\begin{figure}
\bc
\hspace{-1.cm}
\resizebox{9cm}{!}{\includegraphics[]{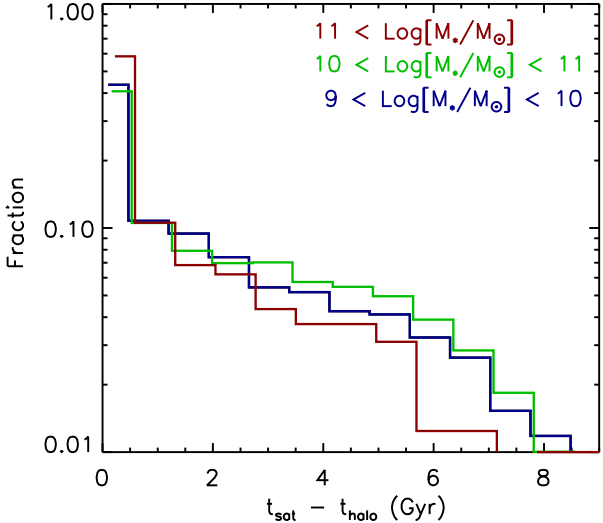}} 
\caption{Distribution of the difference between the lookback time corresponding
  to the event of being accreted onto another system (i.e. becoming a
  satellite), and that corresponding to the accretion onto the final
  group/cluster (see text for details). Different colours correspond to
  galaxies of different present day stellar mass, as indicated in the legend.}
\label{fig:deltaaccr}
\ec
\end{figure}

By construction, the difference between $t_{\rm sat}$ and $t_{\rm halo}$ has to
be positive, or zero.  Figure~\ref{fig:deltaaccr} shows the distribution of
$t_{\rm sat} - t_{\rm halo}$, for galaxies of different stellar mass.  As the
trends are similar for haloes of different mass, we have considered here all
galaxies in all haloes analysed for this study. The figure shows that for $\sim
40$ to $\sim 60$ per cent of the galaxies, the two events considered happen at
the same time. For the remaining galaxies, these two events can differ by up to
$\sim 8$~Gyr. A larger fraction of the most massive galaxies in our sample are
accreted as centrals and, for these galaxies, the difference between the two
events considered is on average smaller than for less massive galaxies. This
results from the combination of different factors. There is a strong
correlation between the stellar mass of central galaxies and the parent halo
mass, and massive haloes form relatively late in hierarchical cosmogonies. So
the most massive galaxies can only be satellites of very massive haloes: the
main progenitor of the parent halo has to be already quite massive in order to
accrete a second system that is also large enough to contain a massive
galaxy. In addition, massive galaxies that were accreted early on during the
history of the parent halo, will be rapidly dragged closer to the inner regions
by dynamical friction, and eventually merge with the central galaxy of the
final halo. Therefore, these galaxies will not appear in our sample that only
contains satellite galaxies surviving at redshift zero.

\begin{figure*}
\bc
\resizebox{16.5cm}{!}{\includegraphics[]{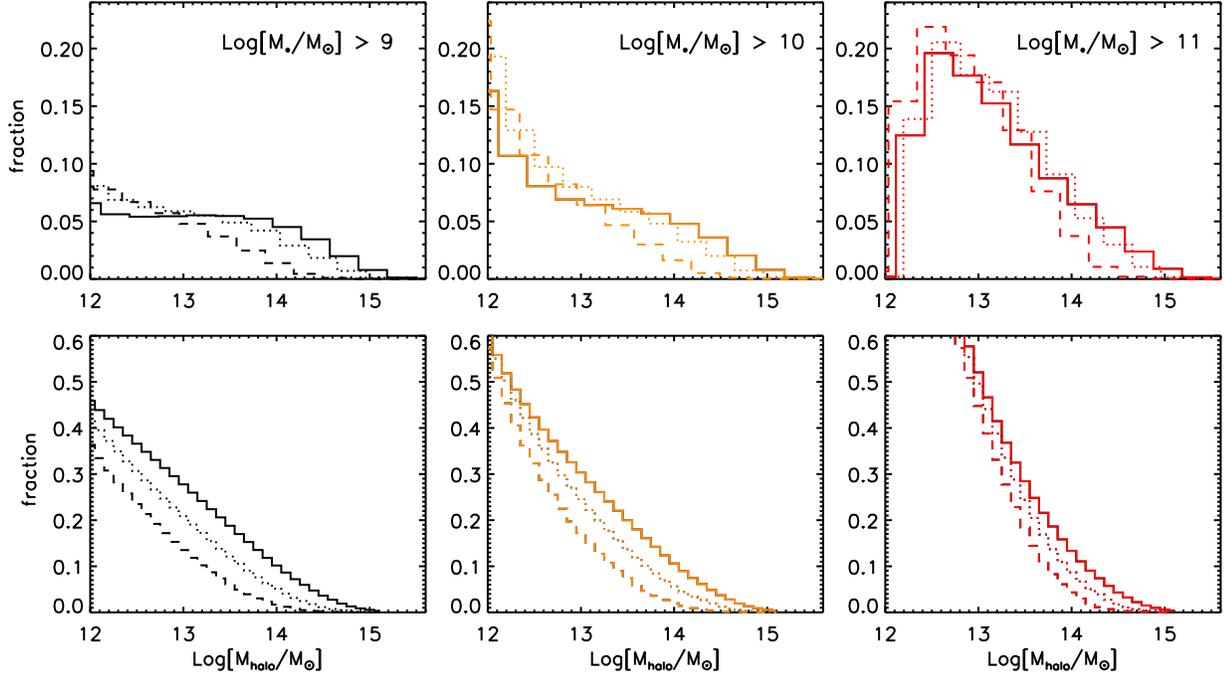}} 
\caption{Differential (top panels) and cumulative (bottom panels) distributions
  of galaxies as a function of their parent halo mass. Different columns
  correspond to different galaxy stellar mass thresholds, as indicated in the
  legend. Different linestyles correspond to different cosmic epochs (solid,
  dotted and dashed lines are used for $z \sim 0$, $0.5$, and $1$,
  respectively). Each distribution is normalized to the total number of
  galaxies above the stellar mass limit considered (without dividing by the bin
  size).}
\label{fig:wherearegx}
\ec
\end{figure*}

Before studying the history of the environments of our model galaxies, it is
interesting to analyse their present day distribution as a function of the
parent halo mass, and how it varies as a function of cosmic
time. Figure~\ref{fig:wherearegx} shows the differential (top panels) and
cumulative (bottom panel) distribution of galaxies for three different galaxy
stellar mass thresholds (increasing from left to right columns). These have
been obtained by using the entire volume of the Millennium Simulation, and all
galaxies with stellar mass $\gtrsim 10^9\,{\rm M_{\odot}}$. Different
linestyles correspond to different cosmic epochs (solid is for present day,
dotted for $z\sim 0.5$, and dashed for $z\sim 1$). Each distribution has been
normalized to the total number of galaxies above the stellar mass limit
indicated in the legend. Note that the cumulative distributions in the bottom
left and middle panels do not reach unity at the lowest halo mass limit shown:
$\sim 55$ per cent of the galaxies with stellar mass larger $\gtrsim 10^9\,{\rm
  M}_{\odot}$ reside today in haloes with mass $< 10^{12}\,{\rm
  M}_{\odot}$. For galaxies more massive than $\sim 10^{10}\,{\rm M}_{\odot}$,
the corresponding fraction drops to $\sim 40$ per cent, and it becomes zero for
galaxies more massive than $\sim 10^{11}\,{\rm M}_{\odot}$.

If one defines as {\it clusters} all haloes more massive than $\sim
10^{14}\,{\rm M}_{\odot}$, the figure shows that only about 10 per cent of the
cosmic galaxy population resides in clusters at present, and that this is
approximately independent of the galaxy stellar mass threshold considered. As
expected, this fraction decreases at higher redshift, as massive clusters form
relatively late. The figure also shows that more massive galaxies tend to
reside in more massive haloes: about 50 per cent of the galaxies with stellar
mass $\gtrsim 10^{11}\,{\rm M}_{\odot}$ reside in haloes more massive than
$\gtrsim 10^{13}\,{\rm M}_{\odot}$, that are sometimes referred to as {\it
  groups} in the literature\footnote{Note that the Local Group, with an
  estimated mass of $5\times 10^{12}\,{\rm M}_{\odot}$
  \citep{Li_and_White_2008}, would not be classified as a group according to
  this definition.}. When considering all galaxies with stellar mass $\gtrsim
10^{10}\,{\rm M}_{\odot}$, however, this fraction drops to only about 30 per
cent at the present day. It is even lower for lower stellar mass
limits. Therefore, it is not generally true that `most galaxies reside in
groups', if by groups one refers to haloes with mass $\gtrsim 10^{13}\,{\rm
  M}_{\odot}$. As just discussed, this statement depends significantly on the
stellar mass limit of the observed sample. We stress that the definitions
adopted in this paper are theoretical ones, and that they do not necessarily
always represent a good proxy for observational definitions. In this paper, we
will work in this `theory space', and use the mass of the parent halo
(specifically, $M_{200}$) as a proxy for the environment. In future work, we
plan to extend our analysis beyond the virial radius, and to use environmental
definitions that are closer to those commonly adopted in the literature.

\section{The history of group and cluster galaxies}
\label{sec:history}

\begin{figure}
\bc
\resizebox{8.4cm}{!}{\includegraphics[]{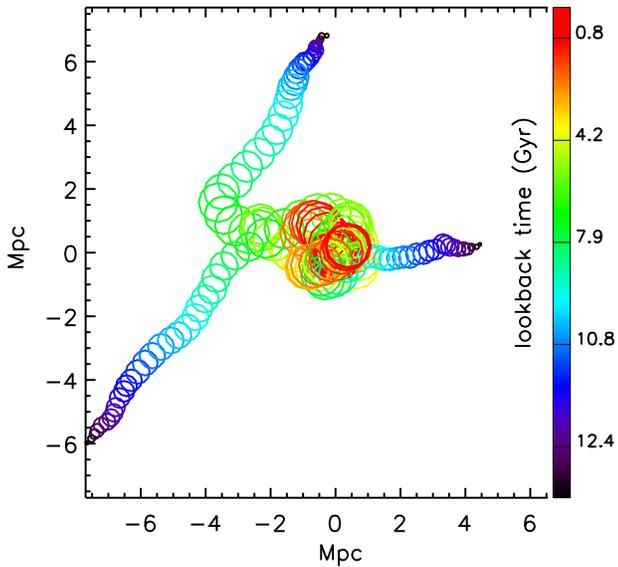}} 
\caption{Projected (comoving) positions of the main progenitors of three
  cluster galaxies, selected randomly between the satellites with stellar mass
  $\gtrsim 10^{11}\,{\rm M}_{\odot}$, residing in haloes with mass $\sim
  10^{14}\,{\rm M}_{\odot}$. Coordinates are re-centred with respect to the
  positions of the main progenitor of the final cluster. The size of the
  symbols scales with the parent halo mass, while the colours encode the
  lookback time.}
\label{fig:galjourney1}
\ec
\end{figure}

\begin{figure}
\bc
\resizebox{8.4cm}{!}{\includegraphics[]{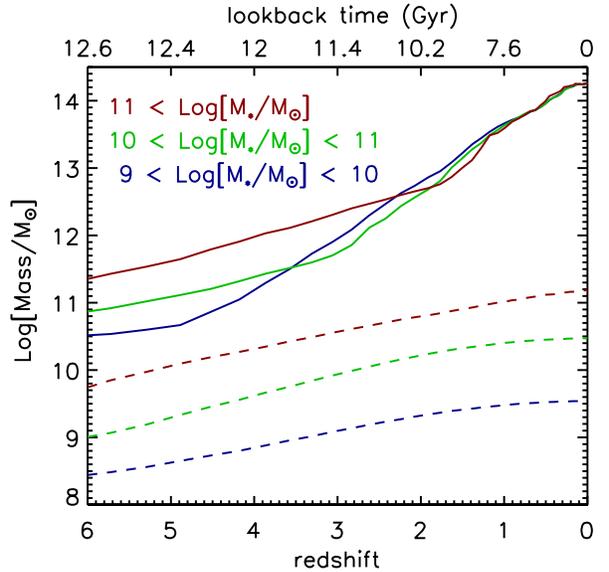}} 
\caption{Mean evolution of the galaxy stellar mass (dashed lines), and of the
  parent halo mass (solid lines). These have been computed by averaging the
  evolution of all satellites of haloes with mass $\sim 10^{14}\,{\rm
    M}_{\odot}$. Different colours correspond to galaxies of different present
  day stellar mass, as indicated in the legend.}
\label{fig:galjourney2}
\ec
\end{figure}

In order to study the environmental history of our model satellite galaxies, we
have traced back their histories and stored the values of their parent halo
mass at all times. In addition, for each galaxy, we have computed the time when
it becomes a satellite, and the time when it is accreted onto the main
progenitor of the final halo, and stored the values of the parent halo mass
corresponding at these times. In the following, we will refer to the value of
the parent halo mass at $t_{\rm sat}$ as ${\rm M}_{\rm halo}({\rm sat})$, and
to the corresponding value at $t_{\rm halo}$ as ${\rm M}_{\rm halo}({\rm
  halo})$.

Figure~\ref{fig:galjourney1} shows the projected comoving trajectories of three
cluster galaxies, re-centred with respect to the positions of the main
progenitor of the final halo at each time. The galaxies have been selected
randomly between the satellites of haloes with mass $\sim 10^{14}\,{\rm
  M}_{\odot}$, and with present stellar mass $\gtrsim 10^{11}\,{\rm
  M}_{\odot}$. The progenitors of these galaxies come from extended regions
around the main progenitor of the final halo, and their parent halo mass
(encoded in the size of the symbols) increases progressively as the galaxies
get closer. Figure~\ref{fig:galjourney2} shows the mean evolution of the
stellar mass (dashed lines) and parent halo mass (solid lines), computed
considering all satellites of 15 haloes with mass $\sim 10^{14}\,{\rm
  M}_{\odot}$ (qualitatively, these results do not change significantly when
considering different ranges of parent halo mass). The mean evolution of the
galaxy stellar mass is very similar in the three mass bins considered. The
evolution of the parent halo masses (solid lines) shows clearly that more
massive galaxies are sitting on average in more massive haloes for most of
their evolution (before starting being accreted onto the final haloes). The
difference in mass is, however, not large. In addition,
Figure~\ref{fig:galjourney2} also shows that the most massive galaxies
surviving at redshift zero fell onto the main progenitor of the final cluster
later than their less massive counterparts: all lines superimpose when galaxies
are in the same halo, and for the most massive galaxies this happens only at
$z\sim 1$.

\begin{figure*}
\bc
\resizebox{16.5cm}{!}{\includegraphics[]{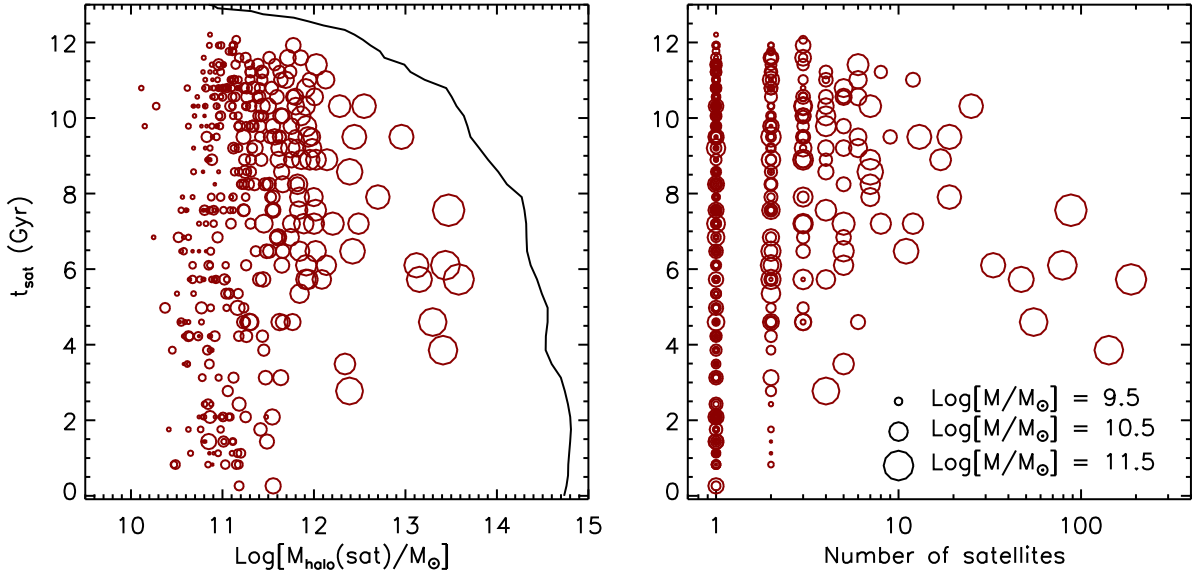}} 
\caption{Left panel: lookback time corresponding to the event of becoming a
  satellite galaxy, as a function of the parent halo mass at the same
  time. Each symbol corresponds to a galaxy, with the symbol size scaling
  proportionally to the present day stellar mass. The example shown corresponds
  to a single halo of mass $M \sim 5\times10^{14}\,{\rm M}_{\odot}$, selected
  at $z=0$. The solid thick line shows the mass evolution of the cluster main
  progenitor. Right: lookback time corresponding to the event of becoming a
  satellite as a function of the galaxy's number of satellites at the same
  time.}
\label{fig:accrvaluesexample}
\ec
\end{figure*}

It is very instructive to look at some properties of group/cluster satellites
at the characteristic times defined above. The left panel of
Figure~\ref{fig:accrvaluesexample} shows the time since the galaxy became a
satellite as a function of the corresponding parent halo mass, for all galaxies
residing in a single cluster of mass $M \sim 5\times10^{14}\,{\rm M}_{\odot}$
at $z=0$. The solid line shows the mass accretion history of the parent
cluster, while the size of the symbols scales with the present day galaxy
stellar mass, as indicated in the legend in the right panel.  The figure shows
that cluster members with low stellar mass become satellites (are accreted onto
more massive haloes) at all times, while the (few) most massive galaxies that
survive as satellites at redshift zero tend to become satellites later compared
to the overall cluster galaxy population. As discussed in the previous section,
this is expected in a hierarchical cosmology.

Since the number of galaxies in a halo increases with halo mass, those galaxies
that became satellites when sitting in massive systems also have larger numbers
of companion galaxies at that time, as shown in the right panel of
Figure~\ref{fig:accrvaluesexample}. Interestingly, the bottom right corners of
both panels in Figure~\ref{fig:accrvaluesexample} appear to be rather `empty'.
The mass of the cluster considered in our particular example grows
significantly down to a lookback times $\sim 4$~Gyr (that corresponds to $z
\sim 0.4$), through the accretion of few relatively massive structures. At
lower redshift, the mass growth of the parent halo is not significant, and it
is mainly driven by the accretion of small haloes and diffuse material. Most of
the galaxies that become satellites at lookback times lower than $\sim 4$~Gyr
are accreted directly onto the main progenitor of the final cluster, and have
low stellar mass. As we will see below, these results hold for the statistical
sample considered in this study. It is interesting that this regime of `slow
mass growth' for the parent haloes coincides with the cosmic epoch that
witnesses the most striking transformations for the cluster galaxy population,
both in terms of their star formation activity and morphological mix
\citep[e.g.][]{Butcher_and_Oemler_1984,Desai_etal_2007}.

\begin{figure*}
\bc
\resizebox{18cm}{!}{\includegraphics[]{figures/distr_accr_values}} 
\caption{Distributions of the times when group/cluster members become satellite
  galaxies (top left panel), and of the times when the galaxies are accreted
  onto the main progenitor of the final halo (bottom left panel). The right
  panels show the corresponding distributions of parent halo masses. The
  distributions shown have been computed using all haloes selected at
  $z=0$. Different colours correspond to different present day galaxy stellar
  mass, as indicated in the legend.}
\label{fig:distraccrvalues}
\ec
\end{figure*}

Figure~\ref{fig:distraccrvalues} shows the distributions of the times when
group/cluster members become satellite galaxies (top left panel), of the times
when the galaxies are accreted onto the final halo (bottom left panel), and the
corresponding distributions of parent halo mass (right panels). As there is no
significant trend as a function of final halo mass, we have stacked together
all haloes in our sample. The top left panel of
Figure~\ref{fig:distraccrvalues} shows that cluster/group members became
satellite galaxies over a wide range of lookback times. The distributions for
low and intermediate mass galaxies exhibit a peak at very early epochs ($\sim
10$~Gyr), while the distribution obtained for the most massive galaxies is
peaked at later times, with most of them becoming satellite between $4$ and
$9$~Gyr ago.

The top right panel of Figure~\ref{fig:distraccrvalues} shows a strong
correlation (albeit with a relatively large scatter) between the present day
galaxy stellar mass and the parent halo mass at $t_{\rm sat}$. This reflects
the strong correlation between halo mass and galaxy stellar mass for central
galaxies. This correlation is preserved for satellite galaxies because their
stellar mass does not increase significantly after being accreted, as their
star formation is efficiently suppressed over a quite short time-scale in our
model.  The distribution of parent halo masses found for the most massive
galaxies is also the widest, due to the fact that the relation between the
galaxy stellar mass and the parent halo mass flattens at high stellar
masses. The distributions shown in the top panels of
Figure~\ref{fig:distraccrvalues} look different when one considers the time
when the galaxies were accreted onto the most massive progenitor of the final
halo ($t_{\rm halo}$). These distributions are shown in the bottom panels of
Figure~\ref{fig:distraccrvalues}. Since, as discussed above, the two times
considered tend to coincide for the most massive galaxies, the distributions
computed for these galaxies are not significantly different. Results are
instead different for low and intermediate stellar mass galaxies. A large
fraction of these tend to be accreted when they are already satellite galaxies,
so that the distributions of $t_{\rm halo}$ are shifted towards later cosmic
epochs with respect to the corresponding distributions of $t_{\rm sat}$. As a
consequence, if one looks at the distribution of halo masses at $t_{\rm halo}$,
this is found to be `bimodal' for these galaxies. The peaks at lower halo
masses are made up of galaxies that are accreted when they are central galaxies
of their own haloes, while the peaks corresponding to larger parent halo masses
are populated by galaxies that are accreted as satellites of relatively large
systems. In particular, we find that $\sim 48$ per cent of the galaxies in the
lowest stellar mass bin considered were accreted as satellite galaxies. The
fraction is slightly lower ($\sim 43$ per cent) for the intermediate mass bin
considered, while it drops to only $\sim 23$ per cent for the most massive
galaxies in our sample.

\begin{figure}
\bc
\hspace{-1.cm}
\resizebox{9cm}{!}{\includegraphics[]{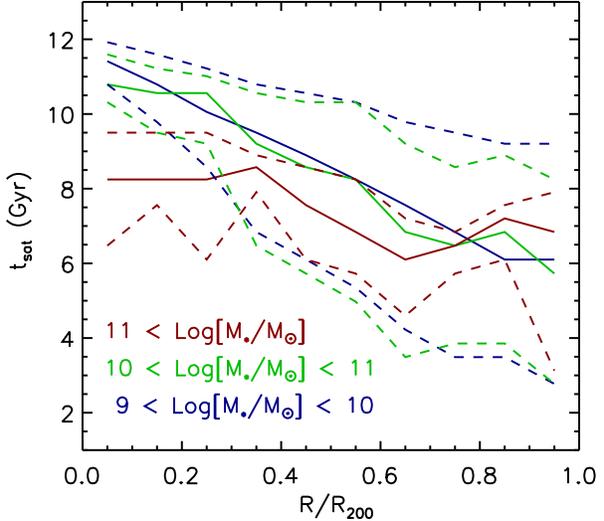}} 
\caption{Median (solid lines) and percentiles (25th and 75th) of the lookback
  times since cluster galaxies became satellites, as a function of their
  present day (three-dimensional) distance from the cluster centre. This has
  been obtained by stacking all haloes with mass $\sim 5\times 10^{14}\,{\rm
    M}_{\odot}$ in our sample. Different colours refer to different galaxy
  stellar mass bins, as indicated in the legend.}
\label{fig:accrtimedist}
\ec
\end{figure}

Substructures that were accreted onto the main progenitor of their parent halo
at early times had relatively short orbital periods. So these should be
located, on average, in the inner regions of the final halo. In addition, these
haloes will have suffered from dynamical friction for a longer time with
respect to haloes of similar mass but accreted later. These two factors combine
to create a strong correlation between the accretion time and the
cluster-centric distance of dark matter substructures
\citep{Gao_etal_2004}. Figure~\ref{fig:accrtimedist} shows that this
correlation is also found for cluster galaxies, although with a large
scatter. Therefore, a radial dependence of galaxy properties is, at least in
part, a natural consequence of the fact that mixing of the galaxy population is
incomplete during cluster assembly. The weakest trend is found for the most
massive satellites surviving at redshift zero that, as discussed above, were
accreted later than their less massive counterparts. The radial trend is
stronger for the low and intermediate stellar mass galaxies considered: those
that are located closer to the centre have been orbiting as satellite galaxies
for about 11~Gyr, while those that are located at the outskirts of the haloes
have been satellites for about 6~Gyr on average. As we will discuss in more
detail in the following section, the trends shown in
Figure~\ref{fig:accrtimedist} can be combined with the observed fractions of
red/passive galaxies as a function of cluster-centric distance to put
constraints on the physical processes (and corresponding timescales)
responsible for the suppression of star formation activity in group and cluster
galaxies. 

A figure similar to our Figure~\ref{fig:accrtimedist} has been shown in
\citet*{Weinmann_vandenBosch_Pasquali_2011} who took advantage of the same
simulated galaxy catalogues but used a cluster sample selected mimicking a
specific observational data-set, and plotted results as a function of the
projected cluster-centric radius. Similar findings have also been discussed in
a recent work by \citet{Smith_etal_2012}. These authors have based their
analysis on publicly available catalogues from the model presented in
\citet[][also based on the Millennium Simulation]{Font_etal_2008}. In
particular, they have selected all galaxies residing in the five most massive
haloes in the simulation, and studied how various characteristic epochs of the
evolution of model galaxies relate to their present day projected distance from
the cluster centre. The results by \cite{Smith_etal_2012} are in very good
agreement with those by \citet{Weinmann_vandenBosch_Pasquali_2011} and with
ours, confirming that these findings are not significantly dependent on the
particular model used, but mainly driven by structure formation.

\section{History bias}
\label{sec:bias}

The results shown in the previous section demonstrate that the galaxy
population of groups and clusters is characterized by a variety of accretion
histories, and that these are strongly dependent on the final galaxy stellar
mass. In this section, we will discuss how the results found above can
influence the observed environmental trends. In particular, we will focus on
two observables: the fraction of passive/red galaxies that reside in haloes of
different mass, and its variation as a function of the cluster-centric
distance.

\begin{figure*}
\bc
\resizebox{18cm}{!}{\includegraphics[]{figures/time_values}} 
\caption{Fraction of galaxies that spent a time longer than that given on the
  x-axis in haloes with mass larger than $10^{12}$ (blue), $10^{13}$ (green),
  and $10^{14}\,{\rm M}_{\odot}$ (orange). Each solid thin line corresponds to
  a single halo, while the corresponding thick lines represent the mean of the
  distributions. Different panels correspond to galaxies residing in haloes of
  different mass today (increasing from left to right), as indicated in the
  legend.}
\label{fig:timevalues}
\ec
\end{figure*}

Figure~\ref{fig:timevalues} shows the fraction of galaxies that spent a time
longer than that given on the x-axis in haloes with mass larger than $10^{12}$
(cyan), $10^{13}$ (green), and $10^{14}\,{\rm M}_{\odot}$ (orange). In this
Figure, each solid thin line corresponds to a different halo, while the thick
solid lines show the mean obtained for the 15 haloes considered in each
sample. As a consequence of the hierarchical structure formation, the fraction
of galaxies that have resided in haloes more massive than $M_{\rm halo}$ for a
time longer than $T_{\rm halo}$ decreases when considering both more massive
haloes and longer times. So it is not surprising that the thick green lines are
always below the thick blue lines, and that these are always above the red
lines. Interestingly, the figure shows a relatively large halo-to-halo scatter:
e.g. focusing on the most massive haloes in our sample, the fraction of
galaxies that have spent more than $6$~Gyr in haloes more massive than
$10^{14}\,{\rm M}_{\odot}$ varies between $\sim 20$ per cent and $\sim 80$ per
cent. If the time spent in a more massive halo can be related to the star
formation activity (or lack thereof) of a galaxy, this halo-to-halo scatter can
be linked to the different fractions of passive galaxies measured in different
clusters \citep[e.g. in the nearby Coma and Virgo
  clusters,][]{Weinmann_etal_2011} and, more in general, to the scatter
measured for some properties of the galaxy populations of groups and clusters
\citep*[see e.g.][]{DeLucia_etal_2012}. Interestingly, the observed fraction of
red/passive galaxies does indeed show a large halo-to-halo scatter, that
appears to increase at lower halo masses
\citep{Poggianti_etal_2006,Balogh_and_McGee_2010}. It should be noted, however,
that observational uncertainties are usually much larger for lower mass
systems. In addition, as we will show in the next sections, the fraction of
passive galaxies increases only weakly as a function of halo mass, and is
larger than $\sim 0.4$ in haloes of mass $\sim 10^{13}\,{\rm M}_{\odot}$. Taken
at face value, these results favour a value for $M_{\rm halo}$ (i.e. the halo
mass above which star formation stars to decline) lower than $10^{14}\,{\rm
  M}_{\odot}$ that would produce a too low fraction of passive galaxies in
haloes with present day mass $\sim 5\times10^{13}\,{\rm M}_{\odot}$ or lower.

We discuss this in more detail in the following sub-sections, where we
investigate how the trends shown in Figure~\ref{fig:timevalues} depend on
galaxy stellar mass and on the cluster-centric distance, and how the
theoretical expectations compare to the estimated fractions of `passive' or
`red' group/cluster galaxies.

\subsection{Observational data}

To constrain our theoretical models, we take advantage of the group catalogue
by \citet{Yang_etal_2007} complemented with the cluster catalogue by
\citet{vonderLinden_etal_2007}, both based on DR4. Specifically, we use the
sample II of the Yang et al. group catalogue, as described in
\citet{vandenBosch_etal_2008}\footnote{The group catalogue is publicly
  available at: http://www.astro.umass.edu/$\sim$xhyang/Group.html}. We use
only satellite galaxies (as in the theoretical predictions discussed below),
and weight results according to the maximum value out to which they can be
observed, to account for Malmquist bias. Our sample of satellite galaxies based
on the Yang et al. catalogue contains 42363 galaxies residing in haloes with
mass larger than $10^{13}\,{\rm M}_{\odot}$. As for the
\citet{vonderLinden_etal_2007} catalogue, their final sample consists of 625
systems at redshifts between 0.03 and 0.1, with masses between $\sim 10^{12}$
and $\sim 10^{15}\,{\rm M}_{\odot}$. In the following, we will consider only
the $214$ systems more massive than $\sim 10^{14}\,{\rm M}_{\odot}$ from this
catalogue.

In addition, we make use of the data catalogues for SDSS DR7 from MPA/JHU to
obtain up-to-date estimates for stellar masses and specific star formation
rates (SSFRs)\footnote{The data are publicly available at:
  http://www.mpa-garching.mpg.de/SDSS/.}. Stellar masses are estimated using
fits to the photometry, and are statistically in agreement with the estimates
from \citet{Kauffmann_etal_2003}. The SSFRs are based on those published in
\citet{Brinchmann_etal_2004}, but with several modifications regarding the
treatment of dust attenuation and aperture corrections. All details can be
found at the reference webpage.

Following \citet{Weinmann_etal_2006a} and \citet{Kimm_etal_2009}, we have
adopted the following demarcation line between red and blue galaxies:
\begin{equation}
^{0.1}(g-r) = 0.7 - 0.032\,(^{0.1}M_r - 5 \,{\rm log}h + 16.5)
\label{eq:colcut}
\end{equation}
All galaxies with ${\rm log}{\rm (SSFR)}< -11$ are classified as passive. As
shown by \citet{Weinmann_etal_2010}, this cut corresponds roughly to the
location of the minimum in the bimodal distribution of SSFRs in the
observations. For comparison, we have also used alternative estimates of the
SSFRs based on UV data from the GALEX satellite, as described in
\citet{McGee_etal_2011}. This results in a reduction of the sample size by a
factor of about 10, as not all SDSS galaxies have GALEX coverage. The scatter
in the observational measurements given below gets larger, but qualitatively
results do not change. Therefore, we only show our measurements of passive
fractions based on the Brinchmann et al. estimates of the SSFRs.

Finally, the cluster sample is complemented with the datasets for Coma and
Virgo that are described in detail in \citet{Weinmann_etal_2011}. For these
datasets, stellar mass estimates are obtained from fits to the photometry,
using the SDSS g- and r-band filters, according to the fitting formula by
\citet*{Zibetti_etal_2009}.

\subsection{Cluster Radial Dependence}

\begin{figure*}
\bc
\resizebox{18cm}{!}{\includegraphics[]{figures/fred_projdist_9_10}} 
\resizebox{18cm}{!}{\includegraphics[]{figures/fred_projdist_10_11}} 
\caption{Fraction of model galaxies that spent more than 3, 5, 7, and 8 Gyr
  (lines of different colour) in haloes more massive than $10^{12}$ (left
  panel), $10^{13}$ (middle panel), and $10^{14}\,{\rm M}_{\odot}$ (right
  panel). Top panels are for galaxies with stellar mass in the range ${\rm
    log}(M_{\rm star}/{\rm M}_{\odot}) = [9-10]$, while bottom panels are for
  galaxies in the range ${\rm log}(M_{\rm star}/{\rm M}_{\odot}) =
  [10-11]$. Theoretical predictions have been obtained averaging all haloes in
  our sample with mass larger than $10^{14}\,{\rm M}_{\odot}$, and projecting
  cluster members on the xy plane. Red diamonds and blue triangles show the red
  fractions for the Coma and Virgo clusters estimated by
  \citet{Weinmann_etal_2011}. Black and red circles show the red and passive
  fractions estimated using the cluster catalogue by
  \citet{vonderLinden_etal_2007}, respectively. Error bars denote confidence
  interval estimates measured as described in the text.}
\label{fig:timevaluesdist}
\ec
\end{figure*}

Figure~\ref{fig:timevaluesdist} shows the fraction of model galaxies that spent
more than 3, 5, 7, and 8 Gyr (lines of different colour) in haloes more massive
than the thresholds indicated in the legend of each panel (increasing from left
to right). Top and bottom panels are for galaxies with stellar masses in the
ranges ${\rm log}(M_{\rm star}/M_{\odot}) = [9-10]$ and ${\rm log}(M_{\rm
  star}/M_{\odot}) = [10-11]$, respectively. Theoretical predictions have been
obtained averaging all haloes in our sample with mass larger than
$10^{14}\,{\rm M}_{\odot}$, and projecting cluster members on the xy plane. Red
diamonds, blue triangles, and black circles show the fractions of red galaxies
for the Coma cluster, for the Virgo cluster, and for the cluster catalogue by
\citet{vonderLinden_etal_2007}, respectively. Finally, red circles show the
fraction of passive galaxies estimated using the catalogue by von der Linden et
al. as in \citet{Weinmann_etal_2010}. Error bars denote confidence intervals
corresponding to a probability of $68$ per cent (1$\sigma$) from quartiles of
the beta distribution \citep[for details on the method, see ][]{Cameron_2011}.

Interestingly, the red fractions estimated from the cluster sample defined in
\citet{vonderLinden_etal_2007} are in quite good agreement with the red
fractions estimated for the Coma and Virgo clusters. This suggests that neither
contamination by background galaxies (this affects at some level the Coma
cluster catalogue as it includes photometric members), nor incompleteness due
to fiber completeness (which affects the cluster catalogue by von der Linden et
al.) influence our results significantly, at least out to a projected radius of
$\sim 1$~Mpc. Both the observed red and passive fractions exhibit a clear
dependency on stellar mass: more massive galaxies exhibit a shallower radial
trend with respect to their less massive counterparts (compare bottom and top
panels of Figure~\ref{fig:timevaluesdist}), in qualitative agreement with the
trends shown in Figure~\ref{fig:accrtimedist}. The measured fractions of
passive galaxies are generally lower than the fractions of red galaxies,
indicating that a fraction of all star forming galaxies are red because
dusty. The offset between the red and passive fractions depends on galaxy
stellar mass and (albeit weakly) on the projected distance from the cluster
centre. In particular, using the results based on the catalogue by
\citet{vonderLinden_etal_2007}, we find that a fraction varying between $\sim
5$ per cent (at small projected distances) to $\sim 10$ per cent (at the
largest projected distances considered) of the red galaxies are star forming
dusty systems in the stellar mass bin ${\rm log}(M_{\rm star}/M_{\odot}) =
[10-11]$. For the lower mass bin considered, the fraction of star forming dusty
galaxies gets larger, varying between $\sim 20$ per cent close to the cluster
centre, to $\sim 24$ per cent at projected distances of $\sim 1$~Mpc. Our
results contradict those by \citet{Wolf_etal_2009} who analyse the properties
of dusty red galaxies in the A901/2 cluster complex at $z\sim 0.17$, and argue
that dusty star forming galaxies are rare for ${\rm log}(M_{\rm
  star}/M_{\odot}) < 10$ while appearing predominantly in the stellar mass
range of ${\rm log}(M_{\rm star}/M_{\odot}) = [10-11]$. We note, however, that
Wolf et al. defined their population of dusty red galaxies differently than
done in our study. In particular, Wolf et al. defined red dusty galaxies from a
full spectral energy distribution fit to the medium-band photometry of the
COMBO-17 survey. This will be more uncertain for faint galaxies (i.e. for low
mass galaxies), a caveat that applies to the passivity criterion used in our
study as well.

The theoretical predictions shown in Figure~\ref{fig:timevaluesdist} are
shallower when considering lower values of $T_{\rm halo}$ at fixed $M_{\rm
  halo}$ values, and for the largest values of $M_{\rm halo}$ considered (right
panels). Again, this is a consequence of hierarchical structure formation: more
massive haloes are formed later than their less massive counterparts so that
galaxies could spend less time in relatively massive systems. A comparison
between theoretical predictions and observational data shows that different
combinations of $T_{\rm halo}$ and $M_{\rm halo}$ can provide a relatively good
agreement with the observed radial trends of passive galaxies. For example, for
galaxies with stellar mass in the range ${\rm log}(M_{\rm star}/M_{\odot}) =
[9-10]$ (upper panels), the observed trend is very close to that obtained when
considering galaxies that have been sitting in haloes more massive than
$10^{13}\,{\rm M}_{\odot}$ for more than $\sim 8$~Gyr. A qualitatively good
agreement (but with a slope shallower than observed) is obtained when
considering galaxies that have been sitting in haloes more massive than
$10^{14}\,{\rm M}_{\odot}$ for more than $\sim 5-7$~Gyr. For the more massive
bin considered (${\rm log}(M_{\rm star}/M_{\odot}) = [10-11]$), the fraction of
passive galaxies has a significantly shallower radial trend. Qualitatively, it
is very similar to the trend obtained considering the fraction of galaxies that
spent more than $\sim 7$~Gyr in haloes more massive than $10^{12}\,{\rm
  M}_{\odot}$, but also to the trend obtained considering galaxies that spent
more than $\sim 5$~Gyr in haloes more massive than $10^{13}\,{\rm M}_{\odot}$,
or more than $\sim 3$~Gyr in haloes more massive than $10^{14}\,{\rm
  M}_{\odot}$.

\subsection{Halo mass dependence}

\begin{figure*}
\bc
\resizebox{18cm}{!}{\includegraphics[]{figures/fred_col_new}} 
\resizebox{18cm}{!}{\includegraphics[]{figures/fpas_br_new}} 
\caption{Fraction of model galaxies that spent more than 3 (left panel), 5
  (middle panel), and 7 Gyr (right panel) in haloes more massive than $10^{12}$
  (dotted lines), $10^{13}$ (dashed lines), and $10^{14}\,{\rm M}_{\odot}$
  (solid lines). Fractions are plotted as a function of the present day halo
  mass.  Different colours correspond to different galaxy stellar mass bins:
  cyan corresponds to ${\rm log}(M_{\rm star}/M_{\odot}) = [9.5-10]$, green to
  ${\rm log}(M_{\rm star}/M_{\odot}) = [10-10.5]$, orange to ${\rm log}(M_{\rm
    star}/M_{\odot}) = [10.5-11]$, and red to ${\rm log}(M_{\rm
    star}/M_{\odot}) = [11-12]$. Data points with error bars are observational
  estimates of red fractions (top panels) and passive fractions (bottom panels)
  based on the DR4 catalogue by \citet{Yang_etal_2007}. Error bars denote
  confidence intervals from quartiles of the beta distribution.}
\label{fig:timevaluesmass}
\ec
\end{figure*}

Another important and independent constraint to our theoretical predictions is
given by the observed fraction of red or passive galaxies as a function of the
present day halo mass.  Figure~\ref{fig:timevaluesmass} compares the fraction
of model galaxies that spent more than 3, 5, and 7 Gyr (from left to right
panels) in haloes more massive than $10^{12}$, $10^{13}$, and $10^{14}\,{\rm
  M}_{\odot}$ (lines of different style) with observational estimates of red
and passive galaxies (top and bottom panels, respectively). To construct the
datasets shown in this figure, we have used the DR4 group catalogue by
\citet{Yang_etal_2007}, as detailed above.

Figure~\ref{fig:timevaluesmass} shows that both the fraction of red galaxies
(top panels) and that of passive galaxies (bottom panels) tend to increase as a
function of halo mass. In addition, at fixed halo mass, the fraction of
red/passive galaxies is larger for more massive galaxies. The results obtained
are in good agreement with those presented in \citet{Kimm_etal_2009}. As noted
above, the fraction of passive galaxies is generally lower than that of red
galaxies. In particular, we find that about 16 per cent of the galaxies are red
but star forming for the lowest galaxy stellar mass bin considered (${\rm
  log}(M_{\rm star}/M_{\odot}) \sim 9.75$). The fraction reduces to about 5 per
cent for the largest stellar mass bin considered (${\rm log}(M_{\rm
  star}/M_{\odot}) \sim 11.25$): i.e. a larger fraction of red massive galaxies
are truly passive, while for less massive galaxies the fraction of star forming
but dusty (and therefore red) objects increases. This is in agreement with what
discussed about Figure~\ref{fig:timevaluesdist}, and in contrast with results
from \citet{Wolf_etal_2009}.

In our models, more than $90$ per cent of the galaxies with stellar mass larger
than ${\rm log}(M_{\rm star}/M_{\odot}) \sim 9.75$ have spent more than 3~Gyr
in haloes more massive than $10^{12}\,{\rm M}_{\odot}$ (dotted lines in the
left panels of Figure~\ref{fig:timevaluesmass}). When considering longer times
(middle and right panels) the fractions decrease, and one can see some
dependency on the galaxy stellar mass (larger fractions are obtained for more
massive galaxies), in particular for the longest times considered. The
fractions of galaxies that have spent equal time in more massive haloes are
lower, as a natural consequence of hierarchical structure formation. In
particular, our model predicts that only about 60 per cent of the galaxies in
the most massive haloes considered in our sample have spent more than 7~Gyr in
haloes more massive than $10^{13}\,{\rm M}_{\odot}$ (dashed lines in right
panels). The fraction decreases rapidly for less massive haloes. For the most
massive galaxies considered (red symbols), the observed passive fractions are
in quite nice agreement with the estimated fractions of galaxies that spent
more than $5$~Gyr in haloes more massive than $10^{13}\,{\rm M}_{\odot}$. For
less massive galaxies, longer times seem to work better (compare cyan lines and
points in the bottom middle and right panels of
Figure~\ref{fig:timevaluesmass}).

\subsection{Central quenching and Satellite quenching}

\begin{figure*}
\bc
\resizebox{18cm}{!}{\includegraphics[]{figures/fquench_col}} 
\resizebox{18cm}{!}{\includegraphics[]{figures/fquench_br}} 
\caption{Lines show the same theoretical predictions given in
  Figure~\ref{fig:timevaluesmass}. Data points with error bars show the red
  (top panels) and passive (bottom panels) transition fractions, i.e. an
  estimate of the fractions of satellites that have been quenched by the
  environment (see text for details).}
\label{fig:fquench}
\ec
\end{figure*}

Putting together both the estimated fractions of passive galaxies as a function
of halo mass and those as a function of cluster-centric distance, there seems
to be a rough timescale/halo mass combination that gives results in quite good
agreement with observational measurements: $M_{\rm halo} \sim 10^{13}\,{\rm
  M}_{\odot}$ and $T_{\rm halo} \sim 5-7$~Gyr is needed for a galaxy to be
quenched with close to $100$ per cent probability, with a tendency for slightly
longer timescales and slightly larger halo masses for lower stellar mass
galaxies. However, the interpretation of the last two figures shown is not
trivial because one has to consider that there is a continuous evolution of the
halo mass hosting the galaxies that end up in a group/cluster. In addition, it
is necessary to account for galaxies that become passive because of `internal'
physical processes ({\it central-quenching}), rather than for `environmental'
processes ({\it satellite quenching}).

The importance of central-quenching can be estimated by considering the
fraction of red/passive central galaxies. Figure~3 by \citet{Kimm_etal_2009}
shows how this fraction varies as a function of the galaxy stellar
mass. Virtually all central galaxies in the first stellar mass bin considered
in our study, ${\rm log}(M_{\rm star}/M_{\odot}) = [9.5-10]$, are active. For
the larger stellar mass bins that we have considered in
Figure~\ref{fig:timevaluesmass}, the fraction of passive central galaxies
increases to $\sim 10-20$ per cent for the bin ${\rm log}(M_{\rm
  star}/M_{\odot}) = [10-10.5]$, and to $\sim 40-50$ per cent for the bin ${\rm
  log}(M_{\rm star}/M_{\odot}) = [10.5-11]$. For the most massive galaxies
considered, $\sim 70$ per cent of the central galaxies are already passive. We
can therefore distinguish two galaxy stellar mass regimes:
\begin{itemize}
\item[(i)] an {\it satellite quenching dominated} regime: galaxies with
  ${\rm log}(M_{\rm star}/{\rm M}_{\odot}) < 10.5$ had their star formation
  rate mainly suppressed by the environment. Our results show that the
  estimated fractions of passive galaxies in this mass regime are in quite good
  agreement with the expected fraction of galaxies that have spent more than
  $\sim 5-7$~Gyr in haloes more massive than $10^{13}\,{\rm M}_{\odot}$;
\item[(ii)] a {\it central quenching dominated} regime: galaxies with ${\rm
  log}(M_{\rm star}/{\rm M}_{\odot}) > 10.5$ whose star formation rate has been
  reduced primarily by internal physical processes. For these galaxies, the
  estimated passive fractions shown in Figure~\ref{fig:timevaluesmass} favour a
  somewhat lower value of $T_{\rm halo}\sim 5$~Gyr. This could be due to the
  fact that these galaxies have spent some fraction of their lifetime as
  central galaxies.
\end{itemize}

In order to determine the importance of environmental processes in suppressing
the star formation rates of satellite galaxies, we follow
\citet{vandenBosch_etal_2008} and define the following {\it transition
  fraction}:
\begin{equation}
{\rm frac}_{\rm \, env} = \frac{{\rm frac}_{\rm \, red, sat} - {\rm frac}_{\rm
    \, red, cen}} {{\rm frac}_{\rm \, blue, cen}}
\label{eq:trans}
\end{equation}
where ${\rm frac}_{\rm \, red, sat}$ is the fraction of satellite galaxies that
are red, ${\rm frac}_{\rm \, red, cen}$ is the fraction of central galaxies
that are red, and ${\rm frac}_{\rm \, blue, cen}$ is the fraction of central
galaxies that are blue. As discussed by \citet{vandenBosch_etal_2008}, under
the assumption that the present day population of central galaxies is
representative of the progenitors of present-day satellite galaxies of the same
stellar mass, the numerator of Eq.~\ref{eq:trans} gives the fraction of all
satellite galaxies that have undergone a blue-to-red transition after their
accretion. Dividing by the fraction of central galaxies that are blue, one
obtains an estimate of the fraction of galaxies that turned red after they
became satellite galaxies (i.e. because of the environmental quenching). An
equivalent quantity can be defined using the measured fractions of passive and
active satellite and central galaxies. In their study,
\citet{vandenBosch_etal_2008} find that satellite quenching affects roughly
$40$ per cent of all galaxies that are still blue at the time of accretion,
independently of their stellar mass.

Our results are shown in the top panel of Figure~\ref{fig:fquench}, and are
compared to the same theoretical predictions that are given in
Figure~\ref{fig:timevaluesmass}. Top panels show the red transition fractions
(Eq.~\ref{eq:trans}), while bottom panels show the corresponding measurements
for passive fractions. As the errors on the transition fractions are dominated
by the errors on the fraction of red/passive satellites, we have shown here the
same error bars used in Figure~\ref{fig:timevaluesmass}. Both our red and
passive transition fractions increase as a function of halo mass. A similar
trend was recently found by \citet{Peng_etal_2011} who used, however, local
overdensity as a proxy for the environment. The red transition fraction also
shows a clear dependency as a function of the galaxy stellar mass: for the two
lowest mass bins considered, the red transition fractions increase from $\sim
55$ per cent in haloes of mass $\sim 10^{13}\,{\rm M}_{\odot}$ to $\sim 75$ per
cent in the most massive haloes considered in our sample ($\sim
5\times10^{14}\,{\rm M}_{\odot}$). For the most massive galaxies considered, no
galaxy has been affected by environmental quenching in the lowest halo mass in
our sample, while the transition fraction increases to $\sim 60$ per cent in
the most massive haloes considered. The passive transition fractions are offset
low with respect to the corresponding red fractions except for the highest
stellar mass bin, and the dependency on stellar mass is weaker. On average, the
passive transition fractions vary between $\sim 30$ per cent in the lowest mass
haloes in our sample to $\sim 65$ per cent in the most massive haloes. For
haloes more massive than $8\times10^{13}\,{\rm M}_{\odot}$, the estimated
passive transition fractions follow quite nicely the trend expected for
galaxies that spent more than $\sim 7$~Gyr in haloes more massive than
$10^{13}\,{\rm M}_{\odot}$. For lower mass haloes, the figure favours a value
of $T_{\rm halo}$ intermediate between 5 and 7~Gyr.

\section{Discussion}
\label{sec:disc}
 
In the previous sections, we have characterized the {\it environmental history}
of group and cluster galaxies. In particular, we have related the observed
fractions of passive/red galaxies to the fraction of galaxies that have spent
more than a given time ($T_{\rm halo}$) in a halo with mass larger than a given
threshold ($M_{\rm halo}$). In this section, we discuss the main caveats of our
analysis, and the main implications of our findings. 

\subsection{Caveats}

As mentioned in Section~\ref{sec:simsam}, the galaxy formation model used in
this study over-predicts the number of low to intermediate stellar mass
galaxies. In particular, \citet{Guo_etal_2011} showed that the model used here
over-predicts the number of galaxies in the stellar mass bin ${\rm log}(M_{\rm
  star}/M_{\odot}) = [9-10]$ by a factor $\sim 2$. For larger stellar masses,
the model reproduces quite nicely the observed galaxy stellar mass function, as
well as the halo occupation distribution measured for galaxy clusters both in
the local Universe and at higher redshift \citep[][but see also
  \citealt{Liu_etal_2010}]{Poggianti_etal_2010}\footnote{Note that
  \citet{Poggianti_etal_2010} used a magnitude limited sample both in their
  observational and model catalogues. We have verified that their magnitude
  limit corresponds approximately to a stellar mass limit of $\sim
  10^{10}\,{\rm M}_{\odot}$.}. If the excess measured in the stellar mass bin
${\rm log}(M_{\rm star}/M_{\odot}) = [9-10]$ is due to galaxies that should
have been stripped and/or merged with the central galaxy of their own parent
halo, then there is likely an excess of galaxies with long survival times as
satellites. Since we base our analysis on the {\it fraction} of galaxies that
spent more than a given time in haloes more massive than some threshold, our
predictions for this mass range should not be significantly affected.

We stress that our results are valid `on average'. As discussed above, there is
a relatively large halo-to-halo variance that contributes, at least in part, to
the scatter in the observed measurements. In addition, one should note that
environmental effects very likely happen in a `probabilistic way': not all
satellites will continue forming stars for some given time and then
simultaneously shut off. Otherwise, we would not observe a bimodality in the
distribution of specific star formation rates. In reality, there will be
satellite galaxies whose star formation rate will be suppressed on relatively
short time-scales and other satellites that will maintain some level of star
formation activity for longer times. This is probably in part related to the
infalling orbital distribution of the satellite galaxies, as the orbital
parameters will influence their dynamical friction
timescale. \citet{McGee_etal_2009} argued that this should not affect
significantly the results obtained if quenching requires relatively long
time-scales, as those suggested by our study. The argument is based on the
relatively narrow distribution of times that infalling dark matter
substructures take to reach their pericentres from the virial radius
\citep[][see also \citealt{Wetzel_2011}]{Benson_2005}. It should be noted,
however, that the evolution of galaxies infalling onto larger haloes depends on
other factors than their orbits, for example on the initial inclination of the
discs with respect to their orbital planes
\citep[e.g.][]{Villalobos_etal_2012}, and variation in the inter-galactic
medium properties of the host. Detailed numerical simulations are therefore
needed to quantify the probability that a galaxy that has been orbiting within
a halo of given mass has its star formation rate suppressed below some given
threshold. As discussed above, our results suggest that the probability to be
quenched approaches unity for galaxies that have spent $5-7$~Gyr in haloes more
massive than $10^{13}\,{\rm M}_{\odot}$. These results might be useful in
guiding the construction of halo occupation distribution models based on
abundance matching for passive and active galaxies.

In Section~\ref{sec:bias}, we have measured the red and passive transition
fractions following \citet{vandenBosch_etal_2008}, and assumed that the
present-day population of central galaxies can be considered representative of
the progenitors of present day satellite galaxies. As discussed in
\citet{vandenBosch_etal_2008}, this assumption is subject to a number of
caveats. In particular: (i) in reality, one should consider the population of
central galaxies at the average time of accretion of the satellite galaxy
population; and (ii) the stellar mass of satellite galaxies might increase if
they continue forming stars after accretion, and might decrease because of
tidal stripping. \citet{vandenBosch_etal_2008} argue that the cumulative effect
should be small. We stress, however, that satellite galaxies in our model have
been accreted over a wide range of cosmic epochs (see top left panel of
Figure~\ref{fig:distraccrvalues}), and that the observed passive fractions
require relatively long time-scales for the transition from blue/active to
red/passive to occur. Therefore, the systematic errors on the transition
fractions estimated in Section~\ref{sec:history} could be relatively large.

\subsection{What is the role of pre-processing?}

In hierarchical cosmogonies, galaxy clusters are assembled through the
accretion of lower mass haloes. Cluster galaxies might have resided for some
time in `groups', making {\it preprocessing} in these systems a potentially
important phase for their evolution \citep{Zabludoff_and_Mulchaey_1998}. This
has been discussed recently in two different studies that have used results
from $N$-body simulations to analyse the accretion histories of cluster
galaxies \citep{Berrier_etal_2009,McGee_etal_2009}. The conclusions obtained by
these studies are quite different: Berrier et al. estimate that $\sim 10$ per
cent of the galaxies residing today in haloes of mass $\sim 10^{14}\,{\rm
  M}_{\odot}$ were accreted onto the cluster potential when residing in haloes
more massive than $10^{13}\,{\rm M}_{\odot}$ (see their figure 2), and argue
that pre-processing in a group environment is ``of secondary importance for
setting cluster galaxy properties for most clusters''. McGee et al. estimate,
for clusters of similar mass, that about $\sim 28$ per cent of the cluster
galaxies have been accreted from haloes more massive than $10^{13}\,{\rm
  M}_{\odot}$ (see their figure 2). They also show that this fraction increases
to $\sim 45$ per cent for the most massive clusters in their sample, arguing
that pre-processing in group environments ``may be significant''.

The reason for this discrepancy lies mainly in the adoption of two different
events corresponding to {\it accretion onto the
  cluster}. \citet{Berrier_etal_2009} use $N$-body simulations of galaxy
clusters (with resolution similar to that of the Millennium Simulation used in
this study), and assume that all bound substructures whose mass at the time of
accretion is larger than some specific threshold, host a galaxy at their
centre. By construction, the time used in Berrier et al. corresponds to our
$t_{\rm sat}$, i.e. the time when a galaxy first becomes satellite of a larger
halo. \citet{McGee_etal_2009} use an approach that is equivalent to ours but
base their analysis of a different semi-analytic model. Similarly to what we
do, they trace the most massive progenitor of each galaxy back in time and
record the halo mass of this progenitor at the time-step just before ``the
galaxy becomes a member of the final cluster'', which should correspond to our
$t_{\rm halo}$. We note that \citet{McGee_etal_2009} did not distinguish
between galaxies that are accreted onto the final cluster as centrals, and
those that are accreted as satellites. Both studies assume that a cluster
galaxy was `pre-processed' if it spent some (significant) fraction of its
lifetime in a `group' halo with mass $\sim 10^{13}\,{\rm M}_{\odot}$ before
`being accreted onto the cluster', independently of whether the galaxy is
a central or a satellite.

As discussed above, the adoption of different characteristic times can lead to
significantly different results, that also depend on the galaxy stellar mass.
Considering only the haloes with mass $\sim 10^{14}\,{\rm M}_{\odot}$ in our
sample, we find that only $\sim 1$ per cent of the galaxies with stellar mass
larger than $10^9\,{\rm M}_{\odot}$ became satellites when residing in haloes
more massive than $\sim 10^{13}\,{\rm M}_{\odot}$. The fraction increases to
$\sim 4$ per cent when considering galaxies more massive than $\sim
10^{10}\,{\rm M}_{\odot}$, and becomes $\sim 37$ per cent when considering the
most massive galaxies in the sample (those with stellar mass larger than $\sim
10^{11}\,{\rm M}_{\odot}$). Indeed, as we have discussed in
Section~\ref{sec:history}, a larger fraction of these massive galaxies are
accreted directly onto the final cluster as centrals. When considering the time
when the galaxy is accreted onto the main progenitor of the cluster, we find
that for the same haloes $\sim 28$, $\sim 29$, and $\sim 44$ per cent of
galaxies with stellar mass larger than $\sim 10^{9}$, $\sim 10^{10}$ and $\sim
10^{11}\,{\rm M}_{\odot}$ were accreted from haloes more massive than $\sim
10^{13}\,{\rm M}_{\odot}$. Our results are therefore in very good agreement
with those found by \citet{McGee_etal_2009}. An accurate comparison with
results from \citet{Berrier_etal_2009} is complicated by the use of a different
approach that does not trace directly the evolution of the galaxy stellar
mass. Their primary cluster sample contains about 16 galaxies per cluster,
which in our case would correspond to a galaxy stellar mass limit of $\sim
2.6\times10^{10}\,{\rm M}_{\odot}$. Considering the fractions estimated above,
our results are therefore not inconsistent with those obtained by
\citet{Berrier_etal_2009}.

If we define as pre-processed all galaxies that have spent time as satellites
of a lower mass system before becoming part of a cluster\footnote{We stress
  that this definition differs from that adopted in some earlier studies,
  including \citet{Berrier_etal_2009} \citet{McGee_etal_2009}, who did not
  distinguish between physical processes acting only on satellite galaxies and
  processes that can affect all galaxies in a group.}, our results show that
the fraction of galaxies that can be pre-processed in a group-size halo of mass
$\sim 10^{13}\,{\rm M}_{\odot}$ is significant. This fraction is largest for
lowest mass galaxies. As discussed above, while $\sim 44$ per cent of the most
massive galaxies have been accreted from haloes more massive than $\sim
10^{13}\,{\rm M}_{\odot}$, a large fraction of those have been accreted as
centrals. On the other hand, basically all the $\sim 28$ per cent low and
intermediate mass galaxies that have been accreted from haloes more massive
than $\sim 10^{13}\,{\rm M}_{\odot}$ are satellites at the time of accretion.

We stress that these results are natural consequences of hierarchical structure
formation, and that they do not depend significantly on the particular
semi-analytic model that we have used in our study. In order to understand how
this preprocessing can affect the observed morphology-density relation,
morphological mix, etc. it is therefore crucial to study the relevance and
effects of various physical processes, at the scales typical of galaxy groups
(specifically, halo masses $\sim 10^{13}\,{\rm M}_{\odot}$). Published
numerical work on the role of environment on galaxy evolution, however, has so
far focused largely on massive galaxy clusters, with very few numerical studies
devoted to understand the effect of the group environment. Further numerical
work in this direction is clearly needed \citep[see
  e.g.][]{Villalobos_etal_2012}.

\subsection{Environment and time-scale of galaxy transformations}

It has long been known that the local galaxy population consists roughly of two
different types of galaxies: red galaxies with low levels of ongoing star
formation and blue galaxies with active star formation. This {\it bimodality}
is known to extend at least up to $z\sim 1$ or higher
\citep{Bell_etal_2004,Whitaker_etal_2011}, and to depend on the environment
\citep[e.g.][]{Baldry_etal_2006}. The physical origin of the observed bimodal
distribution remains a question to be answered. In particular, it is unclear
what is the characteristic time-scale of the galaxy transformation from
blue/active to red/passive, and if this transformation is associated with a
particular environment.

In our study, we have used the group catalogue by \citet{Yang_etal_2007}
complemented with a cluster catalogue \citep{vonderLinden_etal_2007}, and have
tried to constrain these processes using two observational constraints: how the
fraction of passive galaxies varies as a function of halo mass, and how the
same quantity varies as a function of cluster-centric distance. In order to
estimate the efficiency of satellite quenching, we have used the same
statistics discussed by \citet{vandenBosch_etal_2008}. We have shown that the
fraction of satellite galaxies that have been quenched by the environment
increases as a function of halo mass, from $\sim 30$ per cent in the lowest
mass haloes in our sample to about $\sim 65$ per cent in the most massive
haloes considered. The dependency on stellar mass is not large, and it tends to
decrease for increasing halo mass.

Our calculations show that these observational trends are naturally explained
by the fact that a larger fraction of satellite galaxies have spent long times
in relatively massive haloes. Therefore, a more efficient quenching in more
massive haloes is not required. These considerations give little support to
cluster-specific processes like ram-pressure as the main drivers for the
observed trends. In addition, the comparison with our theoretical predictions
show that satellite galaxies become passive after they have spent a time
$T_{\rm halo} \sim 5-7$~Gyr in haloes more massive than $M_{\rm halo} \sim
10^{13}\,{\rm M}_{\odot}$.

Several previous studies have argued for relatively long timescales for
satellite quenching. E.g. \citet*{Balogh_etal_2000} study the origin of
clustercentric gradients in star formation rates and colours of rich clusters
assuming they are built from the accretion of field galaxies, and argue for a
gradual decline (over a timescale of a few Gyr) of the star formation rate of
cluster galaxies after accretion. \citet{Wang_etal_2007} use a physically
based halo occupation distribution model and the observed distribution of
$4000$~\AA~break strengths to constrain the star formation histories of central
and satellite galaxies. They find that satellite galaxies have declining star
formation rates, with average e-folding times $\tau_{\rm s}\sim 2.5$~Gyr, and
no significant dependency on stellar mass. \citet{Kang_and_vandenBosch_2008}
use a semi-analytic model of galaxy formation and the observed increase in the
red fraction of satellite galaxies as a function of stellar mass. They argue
that model results can be brought into good agreement with the observational
data by decreasing the stripping efficiency of the hot gas reservoirs
associated with galaxies being accreted onto a larger halo. In particular, they
assume a constant stripping rate over a time-scale of $\sim 3$~Gyr. Recently,
\citet*{Wetzel_Tinker_and_Conroy_2011} have used a group catalogue based on
SDSS DR7 and studied the distribution of SSFRs for satellite galaxies and its
dependence on stellar mass, halo mass and halo-centric distance. The
`persistent bimodality' they find for satellite galaxies indicate that star
formation in active satellites continues to evolve (as in active centrals) for
several Gyr. Several other studies have argued for relatively long timescales
for the suppression of the star formation rates in satellite galaxies
\citep{Finn_etal_2008,Weinmann_etal_2009,DeLucia_etal_2009,Simard_etal_2009,vonderLinden_etal_2010,McGee_etal_2011}.

As explained in Section~\ref{sec:simsam}, a gradual stripping of the hot gas
reservoir has been advocated and assumed in many recent studies based on
semi-analytic models \citep[see
  e.g.][]{Font_etal_2008,Weinmann_etal_2010,Guo_etal_2011}. It is, however,
unclear how such a gentle mode of strangulation could support cooling and star
formation in satellite galaxies for several Gyr. More detailed numerical
studies are desirable to understand if the main problem with satellite galaxies
in semi-analytic models is related to a poor treatment of environmental
processes, or if it signals a more fundamental problem with the treatment of
star-formation and feedback for these galaxies.

\section{Conclusions}
\label{sec:concl}

We exploit publicly available catalogues from semi-analytic models to study the
environmental history of group and cluster galaxies. In particular, we focused
on haloes with $M_{200} > 10^{13}\,{\rm M}_{\odot}$, carried out our analysis
within a theoretical framework, using the halo mass ($M_{200}$) as a proxy for
the environment, and considered explicitly the dependency on galaxy stellar
mass. Our analysis highlights a number of natural consequences of structure
formation that are important to consider when interpreting observational
data. In particular, we show that:

\begin{itemize}
  \item On average, the surviving massive satellites within galaxy
    groups/clusters were accreted later than their less massive counterparts
    and they come from more massive haloes. The most massive galaxies tend to
    be accreted onto the main progenitor of their final group/cluster when they
    are central galaxies of their own haloes. Less massive group/cluster
    members become satellites over a wide range of redshifts, and about half of
    them are accreted onto the final group/cluster when they are already
    satellite galaxies.
  \item The mixing of galaxy populations is incomplete during cluster assembly,
    which establishes a correlation between the time a galaxy is accreted onto
    a more massive halo (i.e. it becomes a satellite galaxy) and its distance
    from the final cluster centre. The radial trend is weakest for the most
    massive galaxies because of efficient dynamical friction and the late
    formation times of massive haloes.
\end{itemize}

These trends can be considered as the result of a {\it history bias}, that
represents an integral part of the hierarchical framework. According to the
current paradigm for structure formation, dark matter collapses into haloes in
a bottom-up fashion. Small systems form first and subsequently merge to form
progressively larger systems. As structure grows, galaxies join more and more
massive systems, therefore experiencing a variety of environments during their
lifetime. In this framework, where a galaxy resides today (or at any cosmic
epoch) depends heavily on what its past history was.  Our analysis demonstrates
also that binning galaxies according to their stellar mass does not suffice to
disentangle the role of nature and nurture because galaxies of different
stellar mass have different environmental histories. So, for example, two
galaxies of identical mass at some cosmic epoch can end up having different
stellar masses if one of them falls onto a cluster and the other remains in a
region of average density.

The traditional {\it nature versus nurture debate}, as well as the controversy
regarding the primacy of {\it stellar mass versus environment}, represent then
subtle issues regarding `correlations' and `attribution'. It is, in principle,
possible to separate nature vs nurture if they are correlated but physically
uncoupled. However, our results demonstrate that the two are strongly and
physically connected so that any attempt to separate them is ill posed and
misguided. We stress that this does not mean that studying galaxy properties at
fixed stellar mass but in different environments does not provide important
information, as demonstrated by a number of recent studies, many of which are
referred to in this work. Results, however, should be interpreted with care.

To understand and quantify how hierarchical structure formation affects the
observed environmental trends, we considered two observational constraints: the
dependence of the passive galaxy fraction on halo mass, and cluster-centric
distance. We based our observational measurements on the group catalogue by
\citet{Yang_etal_2007} and the cluster catalogue by
\citet{vonderLinden_etal_2007}. We used up-to-date estimates of the star
formation rate and stellar masses, and complemented the cluster data with
observational data available for the Coma and Virgo cluster presented by
\citet{Weinmann_etal_2011}.

We find that significant fractions of group/cluster galaxies have been accreted
onto their final halo as members of groups with mass $\sim 10^{13}\,{\rm
  M}_{\odot}$. For the particular model used in our study, and considering all
haloes in our sample, we find that this fraction is $\sim 27$ per cent for
galaxies with stellar mass $\sim 10^{9}$ and $\sim 10^{10}\,{\rm M}_{\odot}$,
and increases to $\sim 44$ per cent for galaxies with mass $\sim 10^{11}\,{\rm
  M}_{\odot}$. We also find that $\sim 48$, $\sim 43$, and $\sim 23$ per cent
of the galaxies in the same stellar mass bins are accreted onto the main
progenitor of the final halo as satellite galaxies. Large fractions of group
and cluster galaxies have therefore been `pre-processed' as satellites of
groups with mass $\sim 10^{13}\,{\rm M}_{\odot}$. Since a large fraction of the
most massive galaxies we see in clusters today have been accreted as centrals,
they have been pre-processed the least.

Comparisons with observational data suggest that galaxies become passive only
after they have spent $T_{\rm halo} \sim 5-7$~Gyr in haloes more massive than
$M_{\rm halo} = 10^{13}\,{\rm M}_{\odot}$. The observational trends are
naturally explained by the growth of the large scale structure with no need for
a more efficient quenching in the most massive haloes. It is unclear how
satellite galaxies can sustain significant levels of star formation for such
long time-scales and if this can be achieved, within the framework of current
semi-analytic models of galaxy formation, by simply relaxing the assumption of
instantaneous stripping of the hot gas reservoir associated with galaxies when
they are accreted. On the numerical side, very little work has been carried out
to understand the importance and effects of different physical processes at the
velocity dispersions typical of the group environment. Further studies in this
directions are needed.

In this study, we have limited the analysis to haloes selected at redshift
zero.  A more comprehensive analysis extending to higher redshifts is clearly
desirable and will likely provide important constraints on the physical
processes (and related time-scales) responsible for establishing the observed
environmental trends. Finally, we stress that our study is based on theoretical
definitions of environment. In future work, we plan to extend the analysis
outside the virial radius of dark matter haloes, and to use environment
definitions that are closer to those commonly adopted in observational studies
(e.g. local density). We believe that such an analysis will provide important
guidance in the interpretation of observational results in a cosmological
context, and in the comparison of studies at different cosmic epochs and/or
from different surveys.

\section*{Acknowledgements}
The Millennium Simulation databases used in this paper and the web application
providing online access to them were constructed as part of the activities of
the German Astrophysical Virtual Observatory. We are grateful to G. Lemson for
help with the Millennium Database, and to Anna Pasquali for her help with the
SDSS DR7 data. GDL acknowledges financial support from the European Research
Council under the European Community's Seventh Framework Programme
(FP7/2007-2013)/ERC grant agreement n. 202781. We acknowledge fruitful
discussions with Stefano Borgani, Olga Cucciati, Michaela Hirschmann, Anna
Pasquali, Alvaro Villalobos, Dave Wilmann, and Simon White.

\bsp

\label{lastpage}

\bibliographystyle{mn2e}
\bibliography{environmentalhistory}

\end{document}